\documentclass[11pt,fleqn]{article}

\usepackage{amsmath,amssymb,amsthm}

\newcommand{\rank}{\mathop{\rm rank}\nolimits}
\newcommand{\diag}{\mathop{\rm diag}\nolimits}

\newcommand{\noprint}[1]{}

\newtheorem{theorem}{Theorem}

\newtheorem{corollary}{Corollary}

{\theoremstyle{definition} \newtheorem{definition}{Definition}

\begin{document}

\par\noindent {\LARGE\bf
Lowest Dimensional Example on Non-universality\\ of Generalized In\"on\"u--Wigner Contractions\par}

{\vspace{4mm}\par\noindent
Dmytro R. POPOVYCH~$^\dag$ and Roman O. POPOVYCH~$^\ddag$
\par\vspace{2mm}\par}

{\vspace{2mm}\par\it
\noindent $^\dag$Faculty of Mechanics and Mathematics, National Taras Shevchenko University of Kyiv,\\
$\phantom{^{\dag}}$building 7, 2, Academician Glushkov Av., 03127 Kyiv, Ukraine,
\par}

{\vspace{2mm}\par\it
\noindent $^\ddag$Fakult\"at f\"ur Mathematik, Universit\"at Wien, Nordbergstra{\ss}e 15, A-1090 Wien, Austria\\
$\phantom{^{\ddag}}$Institute of Mathematics of NAS of Ukraine, 3 Tereshchenkivska Str., Kyiv-4, Ukraine
 \par}

{\vspace{2mm}\par\noindent $\phantom{^{\dag,\ddag}}$\rm E-mail: \it  deviuss@gmail.com, rop@imath.kiev.ua
 \par}




{\vspace{6mm}\par\noindent\hspace*{5mm}\parbox{150mm}{\small
We prove that there exists just one pair of complex four-dimensional Lie algebras
such that a well-defined contraction among them is not equivalent to a generalized IW-contraction
(or to a one-parametric subgroup degeneration in conventional algebraic terms).
Over the field of real numbers, this pair of algebras is split into two pairs with the same contracted algebra.
The example we constructed demonstrates that even in the dimension four generalized IW-contractions
are not sufficient for realizing all possible contractions,
and this is the lowest dimension in which generalized IW-contractions are not universal. 
Moreover, this is also the first example of nonexistence of generalized IW-contraction
for the case when the contracted algebra is not characteristically nilpotent 
and, therefore, admits nontrivial diagonal derivations.
The lower bound (equal to three) of nonnegative integer parameter exponents which are sufficient
to realize all generalized IW-contractions of four-dimensional Lie algebras is also found.
}\par\vspace{3mm}}

\section{Introduction}

Limiting processes (contractions) of Lie algebras appear in different areas of physics and mathematics,
e.g., in the study of representations, invariants and special functions.
Perhaps the most important example of contraction of Lie algebras arising in
physics is a singular transition from the Poincar\'e algebra to the Galilei one
which corresponds to the limit transition from relativistic mechanics to classical mechanics
when the velocity of light is assumed to go to infinity.
Another important example is the transition from the Heisenberg algebra to the Abelian algebra.
In physical terms the latter means taking the classical limit of quantum mechanics when the Planck constant $\hbar$ goes to zero;
the linear term in the expansion of the commutator in $\hbar$ then yields the Poisson bracket.
It is important to stress that contractions of Lie algebras provide only an initial symmetry background for
limit transitions among physical theories.
Careful analysis of such transitions necessarily includes, in particular,
the study of contractions for representations of Lie algebras associated with these theories.
For example, it was shown in~\cite{LeBellac&LevyLeblond1973} that Maxwell equations admit two possible nonrelativistic limits,
accounting respectively for electric and magnetic effects.
In terms of representations of Lie algebras this means that
the representation of the Poincar\'e algebra corresponding to the Maxwell equations with currents and charges
admits two inequivalent contractions corresponding to the contraction from the Poincar\'e algebra to the Galilei one.
(See also discussion on applications of contractions in~\cite{Nesterenko&Popovych2006} and references therein.)

The concept of the Lie algebra contraction was introduced by Segal~\cite{Segal1951} via limiting processes of bases.
It became well known thanks to the papers by In\"on\"u and Wigner~\cite{Inonu&Wigner1953,Inonu&Wigner1954}
who invented the so-called \emph{In\"on\"u--Wigner contractions} (\emph{IW-contractions}).
A rigorous general definition of contraction, based on limiting processes of Lie brackets, was given by Saletan~\cite{Saletan1961}.
He also studied the entire class of one-parametric contractions
whose matrices are first-order polynomials with respect to contraction parameters.
IW-contractions form a special subclass in the class of \emph{Saletan contractions}.

\looseness=-1
Another extension of the class of IW-contractions was introduced by Doebner and Melsheimer \cite{Doebner&Melsheimer1967}.
They used contraction matrices which become diagonal after choosing suitable bases in the initial and contracted algebras,
with diagonal elements being real powers of the contraction parameters.
(In fact, integer exponents are sufficient, see \cite{Popovych&Popovych2009}
for a simple geometric proof of this longstanding \cite{Weimar-Woods2000} conjecture.)
In the modern physical literature, such contractions are usually called
\emph{generalized In\"on\"u--Wigner contractions}, probably following~\cite{Hegerfeldt1967},
although a number of other names
(\emph{$p$-contractions}, \emph{Doebner--Melsheimer contractions} and \emph{singular IW-contractions}~\cite{Lyhmus1969})
were previously used.
In algebraic papers, similar contractions are called \emph{one-parametric subgroup degenerations}
(in a similar fashion, general contractions are extended to degenerations which are defined for Lie algebras over an arbitrary field
in terms of the orbit closures with respect to the Zariski topology)
\cite{Burde1999,Burde2005,Burde&Steinhoff1999,Grunewald&Halloran1988}.
Note that in fact a one-parametric subgroup degeneration is associated with a one-parametric matrix group only
upon choosing special bases in the corresponding initial and contracted algebras.
Unfortunately, this fact is often ignored.

For a long time it was not known whether any continuous one-parametric contraction can be represented by a generalized IW-contraction.
As all continuous contractions arising in the physical literature enjoy this property,
it was even claimed~\cite{Weimar-Woods2000} that this is true for an arbitrary continuous one-parametric contraction
but the proof presented in~\cite{Weimar-Woods2000} is not correct~\cite{Nesterenko&Popovych2006}.

The first crucial advance in tackling this problem was made in \cite{Burde1999,Burde2005}
where examples of contractions to characteristically nilpotent Lie algebras were constructed
for all dimensions not less that seven.
Since each proper generalized IW-contraction induces a proper grading for the contracted algebra
and each characteristically nilpotent Lie algebra possesses only nilpotent derivations and hence has no proper gradings,
\emph{any contraction to characteristically nilpotent Lie algebras is obviously inequivalent to a generalized IW-contraction}.
Unfortunately, these examples are not yet well known to the physical community.
This is why their detailed analysis and extension to other nilpotent algebras will be a subject of~\cite{Burde&Nesterenko&Popovych2009}.

Contractions of low-dimensional Lie algebras were studied in a number of papers
(see, e.g., \cite{Agaoka1999,Agaoka2002,Burde2005,Burde&Steinhoff1999,Lauret2003,Nesterenko&Popovych2006,Weimar-Woods1991}
and the review of these results in~\cite{Nesterenko&Popovych2006}).
Thus, it was shown in~\cite{Nesterenko&Popovych2006} that
each contraction of complex three-dimensional Lie algebras is equivalent to a simple IW-contraction.
Any contraction of real three-dimensional Lie algebras is realized by
a generalized IW-contraction with nonnegative powers of the contraction parameter which are not greater than two.
Moreover, only the contraction of ${\rm so}(3)$ to the Heisenberg algebra is inequivalent to a simple IW-contraction.
The same result for continuous one-parametric contractions of real three-dimensional Lie algebras was also claimed
in~\cite{Weimar-Woods1991} but contractions within parameterized series of algebras were not explicitly discussed.
All possible contractions of three-dimensional Lie algebras were realized by generalized IW-contractions much earlier
(see, e.g., \cite{Conatser1972,Inonu1964}).
Therefore, the problem was to prove that there are no other contractions of three-dimensional Lie algebras.
For the complex case, it was made in a rigorous way in~\cite{Burde&Steinhoff1999}.

Almost all contractions of four-dimensional Lie algebras were realized in~\cite{Nesterenko&Popovych2006}
via generalized IW-contractions.
For the real case, the exceptions were the contractions
\[
A_{4.10}\to A_{4.1},\quad
2A_{2.1}\to A_{4.1},\quad
2A_{2.1}\to A_1\oplus A_{3.2},\quad
A_{4.10}\to A_1\oplus A_{3.2},
\]
where the above Lie algebras have the following nontrivial commutation relations:
\begin{align*}
2A_{2.1}\colon\quad &[e_1,e_2]=e_1,\ [e_3,e_4]=e_3;\\
A_1\oplus A_{3.2}\colon\quad &[e_2,e_4]=e_2,\ [e_3,e_4]=e_2+e_3;\\
A_{4.1}\colon\quad &[e_2,e_4]=e_1,\ [e_3,e_4]=e_2;\\
A_{4.10}\colon\quad &[e_1,e_3]=e_1,\ [e_2,e_3]=e_2,\ [e_1,e_4]=-e_2,\ [e_2,e_4]=e_1.
\end{align*}
Since the complexifications of the algebras $2A_{2.1}$ and $A_{4.10}$ are isomorphic,
there were only two exceptions in~\cite{Nesterenko&Popovych2006} for the complex case:
$2\mathfrak g_{2.1}\to \mathfrak g_{4.1}$ and $2\mathfrak g_{2.1}\to \mathfrak g_1\oplus \mathfrak g_{3.2}$.
Here $\mathfrak g_{\dots}$ denotes the complexification of the algebra~$A_{\dots}$.

Recently Campoamor-Stursberg found that
in fact both contractions to $A_{4.1}$ are equivalent to generalized IW-contractions \cite{Campoamor-Stursberg2008}.
As remarked by Nesterenko \cite{Nesterenko2008}, the matrix proposed in \cite{Campoamor-Stursberg2008}
for the contraction $2A_{2.1}\to A_{4.1}$ can be optimized via lowering the maximal parameter exponent.
In Section~\ref{SectionOnGenIWcontractionsFrom2G21ToG41} of the present paper 
we first make an algorithmic calculation of an optimized contraction matrix and prove that 
the contraction exponents $(3,2,1,1)$ cannot be lowered for the contraction $2\mathfrak g_{2.1}\to \mathfrak g_{4.1}$ 
and, therefore, the contractions $2A_{2.1}\to A_{4.1}$ and $A_{4.10}\to A_{4.1}$.
The same result for the contraction $\mathrm{so}(3)\oplus A_1\to A_{4.1}$ is obtained 
in Section~\ref{SectionOnGenIWcontractionsFromSO3A1ToA41}. 

In Section~\ref{SectionOnNonexistenceOfGenIWcontractionsFrom2G21ToG1G32} we first provide a detailed proof of the fact
that the contraction $2\mathfrak g_{2.1}\to \mathfrak g_1\oplus \mathfrak g_{3.2}$
is not equivalent to a generalized IW-contraction.
As all other contractions relating complex four-dimensional Lie algebras
were already realized as generalized IW-contractions in \cite{Campoamor-Stursberg2008,Nesterenko&Popovych2006},
we can state the main results of our paper.

\begin{theorem}\label{TheoremOnNonexistenceOfGenIWContractionsBetween4DimComplexLieAlgebras}
There exists a unique contraction among complex four-dimensional Lie algebras
(namely, $2\mathfrak g_{2.1}\to \mathfrak g_1\oplus \mathfrak g_{3.2}$)
which is not equivalent to a generalized In\"on\"u--Wigner contraction.
\end{theorem}

\begin{corollary}\label{CorollaryOnNonexistenceOfGenIWContractionsBetween4DimRealLieAlgebras}
There exist precisely two contractions among real four-dimensional Lie algebras
(namely, $2A_{2.1}\to A_1\oplus A_{3.2}$ and $A_{4.10}\to A_1\oplus A_{3.2}$)
which cannot be realized as generalized In\"on\"u--Wigner contractions.
\end{corollary}

Combining the results of \cite{Campoamor-Stursberg2008,Nesterenko&Popovych2006} with those from
the present paper also yields the following assertion.

\begin{theorem}
Any generalized In\"on\"u--Wigner contraction among complex (resp.\ real) four-dimen\-sional Lie algebras
is equivalent to the one including only nonnegative integer parameter exponents which are not greater than three.
This upper bound is exact. 
The only generalized In\"on\"u--Wigner contractions necessarily involving exponents which do not belong to $\{0,1,2\}$ 
are $2A_{2.1}\to A_{4.1}$, $A_{4.10}\to A_{4.1}$ and $\mathrm{so}(3)\oplus A_1\to A_{4.1}$ in the real case 
and $2\mathfrak g_{2.1}\to\mathfrak g_{4.1}$ in the complex case. 
The minimal tuple of exponents for each of these contractions is $(3,2,1,1)$.
\end{theorem}

The other sections are auxiliary. 
A necessary theoretical background on contractions and generalized IW-contractions is given in 
Section~\ref{SectionOnContractionsGenIWContractionsAndGradings}. 
All components of the technique applied are described in Section~\ref{SectionDiscussionOfTechniqueApplied}.
After summing up results obtained in the paper, in Section~\ref{SectionConclusion} 
open problems of the subject under investigation are formulated.

\section{Contractions, generalized IW-contractions and gradings}
\label{SectionOnContractionsGenIWContractionsAndGradings}

Let $\mathfrak g=(V,[\cdot,\cdot])$ be an $n$-dimensional Lie algebra
with an underlying $n$-dimensional vector space $V$
over $\mathbb R$ or $\mathbb C$ and a Lie bracket~$[\cdot,\cdot]$, $n<\infty$.
Usually a Lie algebra $\mathfrak g$ is defined by
the commutation relations in a~fixed basis~$\{e_1, \ldots, e_n\}$ of $V$.
Namely, it is sufficient to write down the nonzero commutators $[e_i,e_j]=c^{k}_{ij}e_k$, $i<j$,
where $c^{k}_{ij}$ are components of the structure constant tensor of $\mathfrak g$.
In what follows the indices $i$, $j$, $k$, $i'$, $j'$ and $k'$ run from 1 to $n$
and the sum over the repeated indices is understood unless otherwise explicitly stated. 
For a matrix $A$, $a^i_j$ will be the entry of $A$ located at the intersection of the $i$th row and the $j$th column.

Using a continuous mapping $U\colon (0,1]\to {\rm GL}(V)$ we construct a parameterized family of
the Lie algebras $\mathfrak g_{\varepsilon}=(V,[\cdot,\cdot]_{\varepsilon})$, $\varepsilon \in (0,1]$, isomorphic to $\mathfrak g$.
For each $\varepsilon$, the new Lie bracket $[\cdot,\cdot]_{\varepsilon}$ on~$V$ is defined via the old one as follows:
$[x,y]_{\varepsilon}=U_\varepsilon{}^{-1}[U_\varepsilon x,U_\varepsilon y]$ $\forall \; x, y\in V$.

\begin{definition}\label{DefOfContractions1}
If for any $x, y\in V$ there exists the limit
\[
\lim\limits_{\varepsilon \to +0}[x,y]_\varepsilon=
\lim\limits_{\varepsilon \to +0}U_\varepsilon{}^{-1}[U_\varepsilon x,U_\varepsilon y]=:[x,y]_0
\]
then $[\cdot,\cdot]_0$ is a well-defined Lie bracket.
The Lie algebra $\mathfrak g_0=(V,[\cdot,\cdot]_0)$ is called a \emph{one-parametric continuous contraction}
(or just a \emph{contraction}) of the Lie algebra~$\mathfrak g$.
The procedure $\mathfrak g\to\mathfrak g_0$
that yields the Lie algebra~$\mathfrak g_0$ from the algebra~$\mathfrak g$ is also called a \emph{contraction}.
\end{definition}

If a basis of~$V$ is fixed, the operator $U_\varepsilon$ is defined by the corresponding matrix
and Definition~\ref{DefOfContractions1} can be restated in terms of structure constants.
Let ${c}^k_{ij}$ be the structure constants of the algebra~$\mathfrak g$ in the fixed basis~$\{e_1, \ldots, e_n\}$.
Then Definition~\ref{DefOfContractions1} is equivalent to requiring the limit
\[\lim\limits_{\varepsilon\to+0}(U_\varepsilon)_{i'}^i(U_\varepsilon)_{j'}^j(U_\varepsilon{}^{-1})_k^{k'}c^{k}_{ij}=:c^{k'}_{0,i'\!j'}\]
to exist for all values of $i'$, $j'$ and $k'$ and, therefore,
$c^{k'}_{0,i'\!j'}$ are components of the well-defined structure constant tensor of a Lie algebra~$\mathfrak g_0$.
The parameter $\varepsilon$ and the matrix-valued function $U_\varepsilon$ are called a \emph{contraction parameter} and a \emph{contraction matrix},
respectively.

The contraction $\mathfrak g\to\mathfrak g_0$ is called
\emph{trivial} if ${\mathfrak g}_0$ is Abelian and \emph{improper} if ${\mathfrak g}_0$ is isomorphic to ${\mathfrak g}$.

\begin{definition}\label{equivalent contr}
The contractions $\mathfrak g\to\mathfrak g_0$ and $\tilde{\mathfrak g}\to\tilde{\mathfrak g}_0$ are called \emph{(weakly) equivalent}
if the algebras $\tilde{\mathfrak g}$ and $\tilde{\mathfrak g}_0$ are isomorphic to $\mathfrak g$ and $\mathfrak g_0$, respectively.
\end{definition}

Using the weak equivalence concentrates one's attention on existence and results of contractions and
ignores differences in the ways contractions are performed.
To take into account these different ways, we can introduce different notions of stronger equivalence.
Let $\mathop{\rm Aut}(\mathfrak g)$ denote the group of automorphisms of the Lie algebra~$\mathfrak g$.
We identify automorphisms with the corresponding matrices in the canonical basis.

\begin{definition}\label{DefStrongEquivOfContractions}
Two one-parametric contractions in the same pair of Lie algebras $(\mathfrak g,\mathfrak g_0)$
with the contraction matrices $U(\varepsilon)$ and $\tilde U(\varepsilon)$ are called \emph{strongly equivalent} if
there exist $\delta\in(0,1]$, mappings
$\hat U\colon(0,\delta]\to\mathop{\rm Aut}(\mathfrak g)$ and
$\check U\colon(0,\delta]\to\mathop{\rm Aut}(\mathfrak g_0)$
and a continuous monotonic function $\varphi\colon(0,\delta]\to(0,1]$,
$\displaystyle\lim_{\varepsilon\to +0}\varphi(\varepsilon)=0$, such that
$
\tilde U_\varepsilon=\hat U_\varepsilon U_{\varphi(\varepsilon)}\check U_\varepsilon,\quad \varepsilon\in(0,\delta].
$
\end{definition}

The concept of contraction is generalized to arbitrary algebraically closed fields in terms of orbit closures in the variety of Lie algebras
\cite{Burde&Steinhoff1999,Burde1999,Burde2005,Grunewald&Halloran1988,Lauret2003}.
Namely, let $V$ be an $n$-dimensional vector space over an algebraically closed field~$\mathbb F$ and
$\mathcal L_n=\mathcal L_n(\mathbb F)$ denote the set of all possible Lie brackets on~$V$.
We identify $\mu\in\mathcal L_n$ with the corresponding Lie algebra $\mathfrak g=(V,\mu)$.
$\mathcal L_n$ is an algebraic subset of the variety $V^*\otimes V^*\otimes V$ of bilinear maps from $V\times V$ to $V$.
Indeed, upon fixing a basis $\{e_1,\dots,e_n\}$ of~$V$ we have a bijection among $\mathcal L_n$ and
\[
\mathcal C_n=\{(c_{ij}^k)\in\mathbb F^{n^3}\mid c_{ij}^k+c_{ji}^k=0,\,
c_{ij}^{i'\!}c_{i'\!k}^{k'\!}+c_{ki}^{i'\!}c_{i'\!j}^{k'\!}+c_{jk}^{i'\!}c_{i'\!i}^{k'\!}=0\},
\]
which is determined for any Lie bracket $\mu\in\mathcal L_n$ and
any structure constant tuple $(c_{ij}^k)\in\mathcal C_n$ by the formula $\mu(e_i,e_j)=c_{ij}^ke_k$.
$\mathcal L_n$ is called the \emph{variety of $n$-dimensional Lie algebras (over the field~$\mathbb F$)}
or, more precisely, the variety of possible Lie brackets on~$V$.
The group~${\rm GL}(V)$ acts on $\mathcal L_n$ in the following way:
\[
(U\cdot\mu)(x,y)=U\bigl(\mu(U^{-1}x,U^{-1}y)\bigr)\quad \forall U\in {\rm GL}(V),\forall \mu\in\mathcal L_n,\forall x,y\in V.
\]
(It is a left action in contrast to the right action which is more usual for the `physical' contraction tradition
and defined by the formula $(U\cdot\mu)(x,y)=U^{-1}\bigl(\mu(Ux,Uy)\bigr)$.
Of course, this difference is not of fundamental significance.
We use the right action throughout the rest of the paper.)
Denote by $\mathcal O(\mu)$ the orbit of $\mu\in\mathcal L_n$ under the action of~${\rm GL}(V)$ and
by $\overline{\mathcal O(\mu)}$ the closure of $\mathcal O(\mu)$ with respect to the Zariski topology on~$\mathcal L_n$.

\begin{definition}\label{DefOfContractionsViaOrbitClosure}
The Lie algebra $\mathfrak g_0=(V,\mu_0)$ is called a \emph{contraction} (or \emph{degeneration})
of the Lie algebra~$\mathfrak g=(V,\mu)$ if $\mu_0\in\overline{\mathcal O(\mu)}$.
The contraction is \emph{proper} if $\mu_0\in\overline{\mathcal O(\mu)}\backslash\mathcal O(\mu)$.
The contraction is \emph{nontrivial} if $\mu_0\not\equiv0$.
\end{definition}

For $\mathbb F=\mathbb C$ the orbit closures with respect to the Zariski topology coincide
with the orbit closures with respect to the Euclidean topology and Definition~\ref{DefOfContractionsViaOrbitClosure} is
reduced to the usual definition of contractions.

\begin{definition}\label{DefOfGenIWContractions}
The contraction $\mathfrak g\to\mathfrak g_0$ (over $\mathbb C$ or $\mathbb R$)
is called \emph{a generalized In\"on\"u--Wigner contraction} if
its matrix $U_\varepsilon$ can be represented in the form
$U_\varepsilon=AW_\varepsilon P$, where the matrices $A$ and $P$ are nonsingular and constant (i.e., they do not depend on~$\varepsilon$)
and $W_\varepsilon=\diag(\varepsilon^{\alpha_1},\dots,\varepsilon^{\alpha_n})$ for some $\alpha_1,\dots,\alpha_n\in\mathbb R$.
The $n$-tuple of exponents $(\alpha_1,\dots,\alpha_n)$ is called the \emph{signature} of the generalized IW-contraction $\mathfrak g\to\mathfrak g_0$.
\end{definition}

In fact, the signature of a generalized IW-contraction $\mathcal C$ is defined up to a positive multiplier
since the reparametrization $\varepsilon=\tilde\varepsilon^\beta$, where $\beta>0$,
leads to a generalized IW-contraction strongly equivalent to $\mathcal C$.
Moreover, it is sufficient to consider signatures with integer components only.

\begin{theorem}\label{TheoremOnGenIWContractions}
Any generalized IW-contraction is equivalent to
a generalized IW-contraction with an integer signature (and with the same associated constant matrices).
\end{theorem}

This result was believed to hold for a long time (see e.g.\ \cite{Weimar-Woods2000}) but a completely rigorous proof,
which is surprisingly simple, was found only recently in \cite{Popovych&Popovych2009}.

Upon replacing the Lie algebras by isomorphic ones or, in other words,
changing bases in the initial and contracted algebras, we can make the matrices $A$ and $P$ equal to the unit matrix.
This is appropriate for some theoretical considerations but much less so for working with specific Lie algebras.
If $U_\varepsilon=\diag(\varepsilon^{\alpha_1},\dots,\varepsilon^{\alpha_n})$
then the structure constants of the resulting algebra $\mathfrak g_0$ are given by the formula
\[c_{0,ij}^k=\lim_{\varepsilon\to +0}c_{ij}^k\,\varepsilon^{\alpha_i+\alpha_j-\alpha_k}\]
with no sums over the repeated indices.
Therefore, the constraints
$\alpha_i+\alpha_j\geqslant \alpha_k$ if $c_{ij}^k\ne 0$
are necessary and sufficient
for existence of the well-defined generalized IW-contraction with the contraction matrix $U_\varepsilon$,
and $c_{0,ij}^k=c_{ij}^k$ if $\alpha_i+\alpha_j=\alpha_k$ and $c_{0,ij}^k=0$ otherwise.
This obviously implies that the conditions for existence of generalized IW-contractions and the structure of contracted algebras
can be reformulated in the basis-independent fashion in terms of gradings of contracted algebras associated with
filtrations on initial algebras~\cite{Grunewald&Halloran1988,Lyhmus1969}.
(Probably, this was a motivation for introducing the purely algebraic notion of \emph{graded contractions}
\cite{Couture&Patera&Sharp&Winternitz1991,Montigny&Patera1991,Montigny&Patera&Tolar1994,Patera1992}.) In
particular, the contracted algebra~$\mathfrak g_0$ has to admit a derivation
whose matrix is diagonalizable to $\diag(\alpha_1,\dots,\alpha_n)$.%
\footnote{An operator $D\in {\rm GL}(V)$ is a derivation of a Lie algebra~$\mathfrak g$ if $\forall x,y\in V$: $D[x,y]=[Dx,y]+[x,Dy]$. 
The derivations of~$\mathfrak g$ form a Lie algebra ${\rm Der}(\mathfrak g)$ called the derivation algebra of~$\mathfrak g$. 
After the basis $\{e_1,\dots,e_n\}$ of~$V$ fixed, the matrix $\Gamma=(\gamma^i_j)$ is associated with a derivation of~$\mathfrak g$
if and only if its entries satisfy the system 
$
c_{ij}^{k'\!}\gamma^k_{k'\!}=c_{i'\!j}^k\gamma^{i'\!}_i+c_{ij'\!}^k\gamma^{j'\!}_j.
$}

Certain amount of freedom in the matrices~$A$ and~$P$ is preserved even after fixing the canonical commutation relations.
These matrices are defined up to the transformations
\[
\tilde A=MAN, \quad \tilde P=N^{-1}PM_0,
\]
where $M$ and $M_0$ are matrices of  automorphisms for algebras $\mathfrak g$ and $\mathfrak g_0$, respectively,
and $N$ is a matrix commuting with the diagonal part $W_\varepsilon$.
This means that the matrix $N$ corresponds to an
arbitrary change of basis within components of the grading of $\mathfrak g_0$ associated with $W_\varepsilon$.
The generalized IW-contractions with the matrices 
$U_\varepsilon=AW_\varepsilon P$ and 
$\tilde U_\varepsilon=\tilde A\tilde W_\varepsilon \tilde P$,
where 
$\tilde W_\varepsilon=\diag(\varepsilon^{\beta\alpha_1},\dots,\varepsilon^{\beta\alpha_n})$ 
for some $\beta>0$ are obviously equivalent.

Let the canonical basis of $\mathfrak g_0$ be associated with a grading
which is isomorphic to the one induced by the matrix $W_\varepsilon$.
Then the matrix $P$ can be represented as a product $P_{\rm grad}P_{\rm aut}$,
where $P_{\rm grad}$ and $P_{\rm aut}$ are matrices of a change of basis within the graded components
and of an automorphism of~$\mathfrak g_0$, respectively.
Therefore, in such a case we can get rid of the matrix~$P$ by setting it equal to the unit matrix up to the above equivalence.
If there exists a diagonal automorphism in $\mathfrak g_0$ (resp.\ an automorphism in $\mathfrak g$)
with the determinant different from 1 then we
can further set $\det A=1$ in order to simplify the form of the entries of~$A^{-1}$.
Note that such an automorphism indeed exists for all examples we have encountered so far.
If $U_\varepsilon=AW_\varepsilon$, the structure constants of $\mathfrak g_0$ read
\[
c_{0,ij}^k=\lim_{\varepsilon\to +0}a^{i'}_ia^{j'}_jb^k_{k'}c_{i'j'}^{k'}\,\varepsilon^{\alpha_i+\alpha_j-\alpha_k},
\]
where $A=(a^i_j)$, $A^{-1}=(b^i_j)$, and there is no sum over $i$, $j$ and~$k$.

Up to a component permutation, we can introduce a natural ordering within the set of signatures
of generalized IW-contractions with nonnegative integer parameter exponents among two fixed algebras. 
Namely, we assume that $\bar\alpha<\bar\beta$, where  $\alpha_i,\beta_i\in\mathbb Z$,
$\bar\alpha=(\alpha_1,\dots,\alpha_n)$,  $\bar\beta=(\beta_1,\dots,\beta_n)$, 
$\alpha_1\geqslant\dots\geqslant\alpha_n\geqslant0$ and $\beta_1\geqslant\dots\geqslant\beta_n\geqslant0$ 
if $\alpha_1=\beta_1$, \dots, $\alpha_{j-1}=\beta_{j-1}$ and $\alpha_j<\beta_j$ for some~$j$.
An appropriate signature is called \emph{minimal} if it is minimal with respect to the above ordering. 
Finding minimal signatures is often a necessary step for optimizing the contraction matrices.

Simple IW-contractions clearly form a subclass of generalized IW-contractions with signatures equivalent to tuples of zeros and units.
They present limit processes of Lie algebras with contraction matrices of the simplest possible type.
Most contractions of low-dimensional Lie algebras are equivalent to such contractions.
Classifications of IW-contractions for three- and four-dimensional Lie algebras \cite{Conatser1972,Huddleston1978}
can be easily derived from the classifications of subalgebras of such algebras obtained in~\cite{Patera&Winternitz1977}.

\section{Nonexistence of generalized IW-contraction\\ {\mathversion{bold}from $2\mathfrak g_{2.1}$ to $\mathfrak g_1\oplus \mathfrak g_{3.2}$}}
\label{SectionOnNonexistenceOfGenIWcontractionsFrom2G21ToG1G32}

To prove Theorem~\ref{TheoremOnNonexistenceOfGenIWContractionsBetween4DimComplexLieAlgebras}, we use {\em reductio ad absurdum}.
Namely, suppose that the contraction $2\mathfrak g_{2.1}\to \mathfrak g_1\oplus \mathfrak g_{3.2}$ is realized as a generalized IW-contraction.
First of all we should find out which gradings of the algebra $\mathfrak g_1\oplus \mathfrak g_{3.2}$ can be associated with this contraction.

The derivation algebra of $\mathfrak g_1\oplus \mathfrak g_{3.2}$ consists of linear mappings
whose matrices in the canonical basis have the form~\cite{Popovych&Boyko&Nesterenko&Lutfullin2005}
\[
\Gamma=\left(\begin{array}{cccc}
\gamma^1_1&0&0&\gamma^1_4\\[.3ex]
0&\gamma^2_2&\gamma^2_3&\gamma^2_4\\[.3ex]
0&0&\gamma^2_2&\gamma^3_4\\[.3ex]
0&0&0&0
\end{array}\right).
\]
(Recall that the superscript and subscript of a matrix entry denote the corresponding row and column numbers, respectively.)
Therefore, the matrix of any diagonalizable derivation of $\mathfrak g_1\oplus \mathfrak g_{3.2}$ is reduced, by changing the basis,
to the form $\diag(\beta,\alpha,\alpha,0)$, i.e.,
each grading of $\mathfrak g_1\oplus \mathfrak g_{3.2}$ contains a nontrivial component of zero exponent.
In view of this fact, the number of components for any grading associated with
the contraction $2\mathfrak g_{2.1}\to \mathfrak g_1\oplus \mathfrak g_{3.2}$
has to be greater than two because the contraction in question is not equivalent to a simple IW-contraction \cite{Huddleston1978}.
Hence the contraction $2\mathfrak g_{2.1}\to \mathfrak g_1\oplus \mathfrak g_{3.2}$ can generate only gradings
with three nonzero components $L_\beta$, $L_\alpha$ and $L_0$,
where $0\ne\alpha\ne\beta\ne0$, $\dim L_\beta=\dim L_0=1$ and $\dim L_\alpha=2$.
We prove that any such grading~$\tilde{\mathcal G}$ is equivalent, up to automorphisms of $\mathfrak g_1\oplus \mathfrak g_{3.2}$,
to the grading $\mathcal G$ with
$L_\beta=\langle e_1\rangle$, $L_\alpha=\langle e_2,e_3\rangle$ and $L_0=\langle e_4\rangle$.

Indeed, let $\Gamma$ be the matrix (in the canonical basis $\{e_i\}$) of a derivation associated with
a grading $\tilde{\mathcal G}=\{\tilde L_\beta,\tilde L_\alpha,\tilde L_0\}$.
Since the matrix~$\Gamma$ is diagonalizable we have $\gamma^2_3=0$.
We choose a new basis $\tilde e_i=e_js^j_i$, where $|s^i_j|\ne0$, so that
$\tilde L_\beta=\langle \tilde e_1\rangle$, $\tilde L_\alpha=\langle \tilde e_2,\tilde e_3\rangle$ and $L_0=\langle \tilde e_4\rangle$.
Upon this choice the matrix~$\Gamma$ has to be transformed into a diagonal form.
Hence $s^2_1=s^3_1=s^4_1=0$ and $s^1_2=s^4_2=s^1_3=s^4_3=0$.
Then the change of basis in question can be represented as a composition of
the change of basis within the graded components
$\hat e_1=e_1s^1_1$, $\hat e_2=e_2s^2_2+e_3s^3_2$, $\hat e_3=e_2s^2_3+e_3s^3_3$, $\hat e_4=e_4s^4_4$
with $s^1_1s^4_4(s^2_2s^3_3-s^3_2s^2_3)\ne0$,
which does not affect $\Gamma$ in any substantial way,
and of the automorphism $\tilde e_1=\hat e_1$, $\tilde e_2=\hat e_2$, $\tilde e_3=\hat e_3$,
$\tilde e_4=\hat e_4+\hat e_1\hat s^1_4+\hat e_2\hat s^2_4+\hat e_3\hat s^3_4$
setting $\gamma^1_4=\gamma^2_4=\gamma^3_4=0$.
(Here the coefficients $\hat s^1_4$, $\hat s^2_4$ and $\hat s^3_4$ are expressed via $s^i_j$.)
This means that up to the automorphism we can assume
$\tilde L_\beta=L_\beta$, $\tilde L_\alpha=L_\alpha$ and $\tilde L_0=L_0$.

General form of the matrices for the generalized IW-contractions from $2\mathfrak g_{2.1}$ to $\mathfrak g_1\oplus \mathfrak g_{3.2}$
is \mbox{$U_\varepsilon=AW_\varepsilon P$}, where $A$ and $P$ are constant nonsingular matrices and
$W_\varepsilon=\diag(\varepsilon^\beta,\varepsilon^\alpha,\varepsilon^\alpha,1)$.
Since $P$~is a transition matrix among two graded bases with the same signature $(\beta,\alpha,\alpha,0)$,
it admits the representation $P=P_{\rm grad}P_{\rm aut}$,
where $P_{\rm grad}$ and $P_{\rm aut}$ are matrices of change of basis within the graded components
and of an automorphism of~$\mathfrak g_1\oplus\mathfrak g_{3.2}$, respectively.
The matrix~$P_{\rm grad}$ commutes with $W_\varepsilon$ and can be absorbed into the matrix~$A$ by passing from $A$ to $\tilde A=AP_{\rm grad}$.
The matrix~$P_{\rm aut}$ can be ignored as it does not affect the commutation relations of the contracted algebra.
Therefore, it is sufficient to consider only contraction matrices of the form $U_\varepsilon=AW_\varepsilon$ assuming that $P$ is the unit matrix.
Using the scaling automorphisms in $2\mathfrak g_{2.1}$ we can further assume that $\det A=1$.
This assumption significantly simplifies all computations by reducing the size of expressions for the entries of~$A^{-1}$.

Each of the structure constants
$(U_\varepsilon)_{i'}^i(U_\varepsilon)_{j'}^j(U_\varepsilon{}^{-1})_k^{k'}c^{k}_{ij}$ transformed using~$U_\varepsilon$
includes a single power of the parameter~$\varepsilon$.
The set of possible values for the exponents is
\[
\{0,\ \alpha,\ \beta,\ \alpha+\beta,\ \alpha-\beta,\ \beta-\alpha,\ 2\alpha,\ 2\alpha-\beta\}.
\]
We treat the two possible cases $\alpha>\beta$ and $\beta>\alpha$ separately.
In each of these cases we further assume that $\alpha$ and $\beta$ are positive.
Moreover, in the second case we also assume that $2\alpha>\beta$.
The systems of algebraic equations for the entries of the matrix~$A$ derived under the conditions
$\alpha>\beta>0$ or ($\beta>\alpha>0$ and $2\alpha>\beta$) are minimal.
Dropping the additional assumptions leads to extensions of the minimal systems with other algebraic equations. 
Since the minimal systems will be shown to have no solutions, all the extended systems also are incompatible. 
Therefore, it is enough to study only 
the subcase $\alpha>\beta>0$ and the subcase ($\beta>\alpha>0$ and $2\alpha>\beta$) 
of the first and second cases, respectively. 
The parameters~$\alpha$ and~$\beta$ affect only the limiting process $\varepsilon\to0$
and are not explicitly contained in the algebraic equations we have derived. 
The inequalities singling out the subcases completely determine the limit values (either zero or infinity) 
of the above exponents. 
For this reason specific values of the parameters~$\alpha$ and~$\beta$ are not essential.
We can set $\alpha=2$ and $\beta=1$ in the first subcase 
and $\alpha=2$ and $\beta=3$ in the second subcase.

In what follows $B=(b^i_j)$ denotes the inverse $A^{-1}$ of the matrix~$A$.

For the values $\alpha=2$ and $\beta=1$, the conditions for the matrix of the generalized IW-contraction result in the equations
\begin{gather}\label{EqGrading2110a}
\left(\begin{array}{cc}b^1_1&b^1_3\\[.3ex]b^2_1&b^2_3\\[.3ex]b^3_1&b^3_3\end{array}\right)
\left(\begin{array}{c}a^1_1a^2_4-a^2_1a^1_4\\[.3ex]a^3_1a^4_4-a^4_1a^3_4\end{array}\right)=
\left(\begin{array}{c}0\\[.3ex]0\\[.3ex]0\end{array}\right),
\\[.5ex]\label{EqGradingCommon}
\left(\begin{array}{cc}b^2_1&b^2_3\\[.3ex]b^3_1&b^3_3\end{array}\right)Y=
\left(\begin{array}{cc}1&1\\[.3ex]0&1\end{array}\right),
\quad
Y=
\left(\begin{array}{cc}
y^1_1&y^1_2\\[.3ex]
y^2_1&y^2_2
\end{array}\right)
:=
\left(\begin{array}{cc}
a^1_2a^2_4-a^2_2a^1_4&a^1_3a^2_4-a^2_3a^1_4\\[.3ex]
a^3_2a^4_4-a^4_2a^3_4&a^3_3a^4_4-a^4_3a^3_4
\end{array}\right).
\end{gather}
It follows from system~\eqref{EqGradingCommon} that $b^2_1b^3_3-b^2_3b^3_1\ne0$,
$\det Y\ne0$, and hence $(a^1_4,a^2_4)\ne(0,0)$ and $(a^3_4,a^4_4)\ne(0,0)$.
Then $a^1_1a^2_4-a^2_1a^1_4=0$ and $a^1_1a^2_4-a^2_1a^1_4=0$ in view of system~\eqref{EqGrading2110a}, i.e.,
\[
\left(\begin{array}{c}a^1_1\\[.3ex]a^2_1\end{array}\right)=\mu\left(\begin{array}{c}a^1_4\\[.3ex]a^2_4\end{array}\right),
\quad
\left(\begin{array}{c}a^3_1\\[.3ex]a^4_1\end{array}\right)=\mu\left(\begin{array}{c}a^3_4\\[.3ex]a^4_4\end{array}\right).
\]
Since under these conditions we have
\[
\left(\begin{array}{cc}b^2_1&b^2_3\\[.3ex]b^3_1&b^3_3\end{array}\right)=(\nu-\mu)
\left(\begin{array}{rr}a^2_4y^2_2&-a^4_4y^1_2\\[.3ex]-a^2_4y^2_1&a^4_4y^1_1\end{array}\right),
\]
system~\eqref{EqGradingCommon} is expanded into the following set of equations
\begin{gather*}
(\nu-\mu)(a^2_4y^1_1y^2_2-a^4_4y^1_2y^2_1)=1,\quad (\nu-\mu)(a^4_4-a^2_4)y^1_1y^2_1=0,
\\
(\nu-\mu)(a^4_4y^1_1y^2_2-a^2_4y^1_2y^2_1)=1,\quad (\nu-\mu)(a^2_4-a^4_4)y^1_2y^2_2=1.
\end{gather*}
Subtracting the first equation from the third one yields the equation
\[
(\nu-\mu)(a^4_4-a^2_4)(y^1_1y^2_2+y^1_2y^2_1)=0.
\]
As we have $(\nu-\mu)(a^2_4-a^4_4)\ne0$ according to the fourth equation,
system~\eqref{EqGradingCommon} obviously implies
the contradicting conditions $y^1_1y^2_1=0$, $y^1_1y^2_2+y^1_2y^2_1=0$ and $(y^1_1y^2_2,y^1_2y^2_1)\ne(0,0)$.

Therefore, the generalized IW-contraction $2\mathfrak g_{2.1}\to \mathfrak g_1\oplus \mathfrak g_{3.2}$
cannot possess a signature $(\beta,\alpha,\alpha,0)$ with $\alpha>\beta$.

For the values $\alpha=2$ and $\beta=3$ we obtain equations~\eqref{EqGradingCommon} and
\begin{gather}\label{EqGrading3220a}
(a^2_4b^1_1,\,-a^1_4b^1_1,\,a^4_4b^1_3,\,-a^3_4b^1_3)
\left(\begin{array}{ccc}a^1_1&a^1_2&a^1_3\\[.3ex]a^2_1&a^2_2&a^2_3\\[.3ex]a^3_1&a^3_2&a^3_3\\[.3ex]a^4_1&a^4_2&a^4_3\end{array}\right)=
(0,\,0,\,0).
\end{gather}
We attach the identity $a^1_4a^2_4b^1_1-a^2_4a^1_4b^1_1+a^3_4a^4_4b^1_3 -a^4_4a^3_4b^1_3=0$ to
system~\eqref{EqGrading3220a} as the fourth equation.
The extended system can be represented in the form
$(a^2_4b^1_1,-a^1_4b^1_1,a^4_4b^1_3,-a^3_4b^1_3)A=(0,0,0,0)$
and implies, upon multiplying by $B=A^{-1}$ from the right, that
$a^1_4b^1_1=a^2_4b^1_1=a^3_4b^1_3=a^4_4b^1_3=0.$
It follows from system~\eqref{EqGradingCommon} that $\rank Y=2$.
If $a^1_4=a^2_4=0$ (resp. $a^3_4=a^4_4=0$) then $y^1_1=y^1_2=0$ (resp. $y^2_1=y^2_2=0$)
which contradicts the condition $\rank Y=2$.
Therefore, $(a^1_4,a^2_4)\ne(0,0)$, $(a^3_4,a^4_4)\ne(0,0)$ and hence $b^1_1=b^1_3=0$.

In terms of the matrix~$A$, the equations $b^1_1=0$ and $b^1_3=0$ mean that
the minors of~$A$ complementary to~$a^1_1$ and~$a^3_1$ vanish.
Then it follows from the nonsingularity of~$A$ that
the triples $(a^2_2,a^2_3,a^2_4)$ and $(a^4_2,a^4_3,a^4_4)$ are proportional,
and at least one of them has nonzero elements.
In other words, there exist numbers~$\mu$ and~$\nu$, $(\mu,\nu)\ne(0,0)$, and a nonzero triple $(d_2,d_3,d_4)$ such that
$a^2_j=\mu d_j$, $a^4_j=\nu d_j$, $j=2,3,4$.
Upon denoting
\[
\tilde Y=\left(\begin{array}{cc}
\tilde y^1_1&\tilde y^1_2\\[.3ex]
\tilde y^2_1&\tilde y^2_2
\end{array}\right)
:=
\left(\begin{array}{cc}
a^1_2d_4-d_2a^1_4&a^1_3d_4-d_3a^1_4\\[.3ex]
a^3_2d_4-d_2a^3_4&a^3_3d_4-d_3a^3_4
\end{array}\right),
\]
we have
\[
\left(\begin{array}{cc}b^2_1&b^2_3\\[.3ex]b^3_1&b^3_3\end{array}\right)=
(\mu a^4_1-\nu a^2_1)
\left(\begin{array}{rr}\tilde y^2_2&-\tilde y^1_2\\[.3ex]-\tilde y^2_1&\tilde y^1_1\end{array}\right),
\qquad
Y=\left(\begin{array}{rr}\mu\tilde y^1_1&\mu\tilde y^1_2\\[.3ex]\nu\tilde y^2_1&\nu\tilde y^2_2\end{array}\right)
\]
and the matrix equation~\eqref{EqGradingCommon} takes the form
\[
(\mu a^4_1-\nu a^2_1)
\left(\begin{array}{cc}
\mu\tilde y^1_1\tilde y^2_2-\nu\tilde y^2_1\tilde y^1_2&(\mu-\nu)\tilde y^2_2\tilde y^1_2
\\[.3ex]
(\nu-\mu)\tilde y^1_1\tilde y^2_1&\nu\tilde y^1_1\tilde y^2_2-\mu\tilde y^2_1\tilde y^1_2
\end{array}\right)=
\left(\begin{array}{cc}1&1\\[.3ex]0&1\end{array}\right),
\]
We pick the equation for the $(1,2)$-entries and
two combinations of the equations for $(1,1)$- and $(2,2)$-entries with the coefficients $(\mu,-\nu)$ and $(\nu,-\mu)$:
\begin{gather*}
(\mu a^4_1-\nu a^2_1)(\mu-\nu)\tilde y^2_2\tilde y^1_2=1,\\
(\mu a^4_1-\nu a^2_1)(\mu^2-\nu^2)\tilde y^1_1\tilde y^2_2=\mu-\nu,\\
(\mu a^4_1-\nu a^2_1)(\mu^2-\nu^2)\tilde y^2_1\tilde y^1_2=\nu-\mu.
\end{gather*}
These equations imply
$\mu a^4_1-\nu a^2_1\ne0$, $\mu-\nu\ne0$, $\tilde y^1_1\ne0$ and $\tilde y^2_1\ne0$,
and the latter contradict the equation $(\mu a^4_1-\nu a^2_1)(\mu-\nu)\tilde y^1_1\tilde y^2_1=0$ for $(2,1)$-entries.
Therefore, the matrix $U_\varepsilon$ of
the generalized IW-contraction $2\mathfrak g_{2.1}\to \mathfrak g_1\oplus \mathfrak g_{3.2}$
cannot have diagonal part of the form
$W_\varepsilon=\diag(\varepsilon^\beta,\varepsilon^\alpha,\varepsilon^\alpha,1)$ with $\alpha<\beta$.

Since assuming existence of generalized IW-contractions
from $2\mathfrak g_{2.1}$ to $\mathfrak g_1\oplus \mathfrak g_{3.2}$ leads to contradiction for
all possible values of the parameter exponents, this assumption is not true.
Taking into account the results of~\cite{Campoamor-Stursberg2008,Nesterenko&Popovych2006},
we finally arrive at Theorem~\ref{TheoremOnNonexistenceOfGenIWContractionsBetween4DimComplexLieAlgebras}.

The ground field (complex or real) is not essential for the proof.
Therefore, the statement on the contraction among the algebras $2\mathfrak g_{2.1}$ and $\mathfrak g_1\oplus \mathfrak g_{3.2}$
can be directly reformulated for the contraction among their real counterparts
$2A_{2.1}$ and $A_1\oplus A_{3.2}$.
Moreover, if the contraction $A_{4.10}\to A_1\oplus A_{3.2}$ could be realized by a generalized IW-contraction
over~$\mathbb R$ then the same statement would be true over~$\mathbb C$ for its complexification which is equivalent to
the contraction $2\mathfrak g_{2.1}\to\mathfrak g_1\oplus \mathfrak g_{3.2}$.
This contradicts the proved nonexistence of generalized IW-contraction among
$2\mathfrak g_{2.1}$ and $\mathfrak g_1\oplus \mathfrak g_{3.2}$.
As a result, we obtain Corollary~\ref{CorollaryOnNonexistenceOfGenIWContractionsBetween4DimRealLieAlgebras}.

\section{{\mathversion{bold} Generalized IW-contractions from $2\mathfrak g_{2.1}$ to $\mathfrak g_{4.1}$}}
\label{SectionOnGenIWcontractionsFrom2G21ToG41}

In analogy with the study of the contraction $2\mathfrak g_{2.1}\to \mathfrak g_1\oplus \mathfrak g_{3.2}$,
consider first the gradings of the contracted algebra.
The derivation algebra of $\mathfrak g_{4.1}$ is formed by the linear mappings
whose matrices in the canonical basis have the form~\cite{Popovych&Boyko&Nesterenko&Lutfullin2005}
\begin{equation}\label{EqDerMatrixOfG41}
\Gamma=\left(\begin{array}{cccc}
\gamma^3_3+2\gamma^4_4&\gamma^2_3&\gamma^1_3&\gamma^1_4\\[.3ex]
0&\gamma^3_3+\gamma^4_4&\gamma^2_3&\gamma^2_4\\[.3ex]
0&0&\gamma^3_3&\gamma^3_4\\[.3ex]
0&0&0&\gamma^4_4
\end{array}\right).
\end{equation}
Any diagonalizable matrix of the form~\eqref{EqDerMatrixOfG41} can be reduced, upon a suitable change of basis,
to the form $\diag(\alpha+2\beta,\alpha+\beta,\alpha,\beta)$,
where $\alpha=\gamma^3_3$ and $\beta=\gamma^4_4$.
The contraction $2\mathfrak g_{2.1}\to\mathfrak g_{4.1}$ is not equivalent to a simple IW-contraction \cite{Huddleston1978}.
Hence the quadruple with $\alpha=1$ and $\beta=0$ cannot be a signature for this contraction.
We study other quadruples corresponding to minimal nonnegative integer values of~$\alpha$ and~$\beta$,
namely, the quadruples $(4,3,2,1)$, $(3,2,1,1)$, $(2,1,0,1)$.

The first two of these quadruples are signatures of generalized IW-contractions from $2\mathfrak g_{2.1}$ to $\mathfrak g_{4.1}$.
Considering them, we from the very beginning restrict ourselves to looking for the contraction matrices
in the generalized IW-form 
with $P$ equal to the unit matrix and $\det A=1$.

The quadruple $(4,3,2,1)$ leads to a system involving only three equations for entries of
~$A$:
\begin{gather}\label{EqG41Grading4321a}
\left(\begin{array}{cc}b^1_1&b^1_3\\[.3ex]b^2_1&b^2_3\end{array}\right)
\left(\begin{array}{c}a^1_3a^2_4-a^2_3a^1_4\\[.3ex]a^3_3a^4_4-a^4_3a^3_4\end{array}\right)=
\left(\begin{array}{c}0\\[.3ex]1\end{array}\right),
\\[.5ex]\label{EqG41Grading4321b}
(a^1_2a^2_4-a^2_2a^1_4)b^1_1+(a^3_2a^4_4-a^4_2a^3_4)b^1_3=1.
\end{gather}
Recall that $(b^i_j)=A^{-1}$.
A particular solution of system \eqref{EqG41Grading4321a}--\eqref{EqG41Grading4321b} was implicitly found in~\cite{Campoamor-Stursberg2008}.

For the parameter exponents $(3,2,1,1)$ the system \eqref{EqG41Grading4321a}--\eqref{EqG41Grading4321b} is extended with a single equation
\begin{gather}\label{EqG41Grading3211}
(a^1_2a^2_3-a^2_2a^1_3)b^1_1+(a^3_2a^4_3-a^4_2a^3_3)b^1_3=0.
\end{gather}
We obtain a solution of the whole system \eqref{EqG41Grading4321a}--\eqref{EqG41Grading3211} under the constraint~$\det A=1$.
Hence the suggested matrix~$A$ will be admissible for generalized IW-contractions from $2\mathfrak g_{2.1}$ to $\mathfrak g_{4.1}$
with both signatures $(4,3,2,1)$ and $(3,2,1,1)$.
Since system \eqref{EqG41Grading4321a}--\eqref{EqG41Grading3211} is underdetermined,
we can choose simple values for the most of $a^i_j$ without breaking compatibility of the equations that are not satisfied.
It follows from~\eqref{EqG41Grading4321a} and~\eqref{EqG41Grading4321b} that
$(b^1_1,b^1_3)\ne(0,0)$ and $(b^2_1,b^2_3)\ne(0,0)$.
Should we have $b^1_1b^2_3-b^1_3b^2_1=0$, $(b^1_1,b^1_3)$ would equal $\mu(b^2_1,b^2_3)$ for some $\mu\ne0$
and equation~\eqref{EqG41Grading4321a} would imply the contradictory condition $\mu=0$.
Therefore,
\[
b^1_1b^2_3-b^1_3b^2_1=-(a^2_3a^4_4-a^2_4a^4_3)\ne0.
\]
We set $a^2_3=a^4_3=a^4_4=1$ and $a^2_4=0$.
Then $a^1_2=a^3_3=0\bmod\mathop{\rm Aut}(2\mathfrak g_{2.1})$.
After substituting the fixed values of $a$'s, system \eqref{EqG41Grading4321a}--\eqref{EqG41Grading3211} yields, in particular,
$a^1_2a^3_4-a^1_4a^3_2=0$ and $a^1_1a^3_4-a^1_4a^3_1=1$.
For simplicity we also choose $a^1_4=a^3_4=a^1_1=1$ and $a^2_1=a^4_1=a^1_2=a^2_2=0$.
The remaining entries of~$A$ are readily found.
As a result, we obtain the solution
\[
A=\left(\begin{array}{cccc}
1&0&0&1\\0&0&1&0\\0&0&0&1\\0&1&1&1
\end{array}\right).
\]
The matrix $U_\varepsilon=A\diag(\varepsilon^3,\varepsilon^2,\varepsilon,\varepsilon)$, found by us,
realizes a generalized IW-contraction $2\mathfrak g_{2.1}\to\mathfrak g_{4.1}$ and
is simpler than the one presented in~\cite{Campoamor-Stursberg2008}.

Now we prove using {\em reductio ad absurdum}
that the quadruple $(2,1,0,1)$ cannot be a signature of a generalized IW-contraction $2\mathfrak g_{2.1}\to\mathfrak g_{4.1}$.

Indeed, suppose that there {\em exists} a generalized IW-contraction $2\mathfrak g_{2.1}\to\mathfrak g_{4.1}$ with the signature $(2,1,0,1)$.
This means that for some nonsingular constant matrices~$A$ and~$P$ the product
$U_\varepsilon=A\diag(\varepsilon^2,\varepsilon,1,\varepsilon)P$ is a matrix of the contraction $2\mathfrak g_{2.1}\to\mathfrak g_{4.1}$.
The Lie algebra obtained by the contraction with the matrix $A\diag(\varepsilon^2,\varepsilon,1,\varepsilon)$
from the algebra $2\mathfrak g_{2.1}$ possesses the derivation with the matrix $\diag(2,1,0,1)$,
which should be transformed under the action of~$P$ into a matrix~$\Gamma$
of the form~\eqref{EqDerMatrixOfG41} with $\gamma_{33}=0$ and $\gamma_{44}=1$.
Therefore, the matrices~$P$ and~$\Gamma$ satisfy the equation $\diag(2,1,0,1)P=P\Gamma$
which implies the diagonalizability condition $\gamma^1_2\gamma^3_4+\gamma^2_4=0$ for~$\Gamma$
and the representation $P=P_{\rm grad}P_{\rm aut}$, where
\[
P_{\rm grad}=\left(\begin{array}{cccc}
p^1_1&0&0&0\\0&p^2_2&0&p^2_4\\0&0&p^3_3&0\\0&p^4_2&0&p^4_4
\end{array}\right)
\quad\mbox{and}\quad
P_{\rm aut}=\left(\begin{array}{cccc}
1&\gamma^1_2&\sigma_1&\sigma_2\\
0&1&\gamma^1_2&-\gamma^1_2\gamma^3_4\\0&0&1&-\gamma^3_4\\0&0&0&1
\end{array}\right)
\]
are matrices of a change of basis within the graded components
and of an automorphism of~$\mathfrak g_{4.1}$ in the canonical basis, respectively,
$\sigma_1=\frac12(\gamma^1_3+(\gamma^1_2)^2)$ and $\sigma_2=\gamma^1_4+\frac12\gamma^3_4(\gamma^1_3-(\gamma^1_2)^2)$.
Taking into account the representation for~$P$, we can assume $P$ to be equal to the unit matrix and
consider only contraction matrices of the form $U_\varepsilon=AW_\varepsilon$.

In contrast with the two first signatures, the conditions for the matrix of generalized IW-contractions with the signature $(2,1,0,1)$
result in a much larger system consisting of eight equations.
We can represent them in the form
\begin{gather}\label{EqG41Grading2101a}
(a^2_3b^1_1,\,-a^1_3b^1_1,\,a^4_3b^1_3,\,-a^3_3b^1_3)
\left(\begin{array}{ccc}a^1_1&a^1_2&a^1_4\\[.3ex]a^2_1&a^2_2&a^2_4\\[.3ex]a^3_1&a^3_2&a^3_4\\[.3ex]a^4_1&a^4_2&a^4_4\end{array}\right)=
(0,\,0,\,0),
\\[.5ex]\label{EqG41Grading2101b}
\left(\begin{array}{cc}b^1_1&b^1_3\\[.3ex]b^2_1&b^2_3\\[.3ex]b^4_1&b^4_3\end{array}\right)Y=
\left(\begin{array}{cc}0&0\\[.3ex]0&1\\[.3ex]0&0\end{array}\right),
\quad
Y=
\left(\begin{array}{cc}
y^1_1&y^1_2\\[.3ex]
y^2_1&y^2_2
\end{array}\right)
:=
\left(\begin{array}{cc}
a^1_2a^2_3-a^2_2a^1_3&a^1_3a^2_4-a^2_3a^1_4\\[.3ex]
a^3_2a^4_3-a^4_2a^3_3&a^3_3a^4_4-a^4_3a^3_4
\end{array}\right).
\\[1.5ex]\label{EqG41Grading2101c}
(a^1_2a^2_4-a^2_2a^1_4)b^1_1+(a^3_2a^4_4-a^4_2a^3_4)b^1_3=1.
\end{gather}
A pair of equations is included in both~\eqref{EqG41Grading2101a} and~\eqref{EqG41Grading2101b} for convenience.

From~\eqref{EqG41Grading2101b} and~\eqref{EqG41Grading2101c} we infer that $y^1_1=y^2_1=0$.
Indeed, otherwise we would have
\[
\left(\begin{array}{cc}b^1_1\\[.3ex]b^2_1\\[.3ex]b^4_1\end{array}\right)=-y^2_1
\left(\begin{array}{cc}d^1\\[.3ex]d^2\\[.3ex]d^4\end{array}\right),
\quad
\left(\begin{array}{cc}b^1_3\\[.3ex]b^2_3\\[.3ex]b^4_3\end{array}\right)=y^1_1
\left(\begin{array}{cc}d^1\\[.3ex]d^2\\[.3ex]d^4\end{array}\right),
\quad
(y^1_1y^2_2-y^1_2y^2_1)\left(\begin{array}{cc}d^1\\[.3ex]d^2\\[.3ex]d^4\end{array}\right)=
\left(\begin{array}{cc}0\\[.3ex]1\\[.3ex]0\end{array}\right),
\]
i.e., $y^1_1y^2_2-y^1_2y^2_1\ne0$, $d_1=d_4=0$ and, therefore, $b^1_1=b^1_3=0$ which contradicts equation~\eqref{EqG41Grading2101c}.

We attach the identity $a^1_3a^2_3b^1_1 -a^2_3a^1_3b^1_1+a^3_3a^4_3b^1_3 -a^4_3a^3_3b^1_3=0$
to system~\eqref{EqG41Grading2101a} as the fourth equation.
After reordering equations, the extended system can be represented in the form
$(a^2_3b^1_1,-a^1_3b^1_1,a^4_3b^1_3,-a^3_3b^1_3)A=(0,0,0,0)$.
Since $\det A\ne0$, we find that
\[
a^1_3b^1_1=a^2_3b^1_1=a^3_3b^1_3=a^4_3b^1_3=0
\]
and, therefore, $b^1_1b^1_3=0$ in view of $(a^1_3,a^2_3,a^3_3,a^4_3)^{\rm T}\ne(0,0,0,0)^{\rm T}$.
It follows from~\eqref{EqG41Grading2101c} that $(b^1_1,b^1_3)\ne(0,0)$.
This is why there are two possible cases for values $(b^1_1,b^1_3)$, namely,
\[
b^1_1\ne0,\ b^1_3=0\quad\mbox{and}\quad b^1_1=0,\ b^1_3\ne0.
\]

Below we consider the first case only. The second one is treated in a similar fashion.

If $b^1_1\ne0$ and $b^1_3=0$ then $a^1_3=a^2_3=0$ and hence $y^1_2=0$, $b^2_3y^2_2=1$, $b^4_3y^2_2=0$.
This leads to the conditions $b^2_3\ne0$, $y^2_2\ne0$ and $b^4_3=0$.
In terms of the matrix~$A$, vanishing of $b^1_3$ and $b^4_3$ means that
the triples $(a^1_2,a^2_2,a^4_2)$ and $(a^1_3,a^2_3,a^4_3)$ are proportional.
Then $(a^1_2,a^2_2)\ne(0,0)$ and $a^4_3=0$ in view of  $a^1_3=a^2_3=0$ and $\det A\ne0$.
Since $a^1_3=a^2_3=a^4_3=0$, we obtain the equality $b^2_3=0$ contradicting the earlier inequality $b^2_3\ne0$.

As a result, we see that the quadruple $(3,2,1,1)$ is the signature of a generalized IW-contraction $2\mathfrak g_{2.1}\to\mathfrak g_{4.1}$
with minimal nonnegative integer exponents.

The proof of minimality remains valid if we use the real (instead of the complex) numbers  as the ground field. 
Hence the above statement on the contraction among the algebras $2\mathfrak g_{2.1}$ to $\mathfrak g_{4.1}$
can be directly reformulated for the contraction among their real counterparts
$2A_{2.1}$ and $A_{4.1}$.
Moreover, it is known \cite{Campoamor-Stursberg2008}
that the contraction $A_{4.10}\to A_{4.1}$ is realized by a generalized IW-contraction with the signature $(3,2,1,1)$. 
Should this contraction have a signature with components only from $\{0,1,2\}$ over~$\mathbb R$, 
the same statement would be valid over~$\mathbb C$ for its complexification 
which is equivalent to the contraction $2\mathfrak g_{2.1}\to\mathfrak g_{4.1}$.
This contradicts the above assertion on the minimality of the signature $(3,2,1,1)$ in the complex case.
Thus, for both contractions $2A_{2.1}\to A_{4.1}$ and $A_{4.10}\to A_{4.1}$
the quadruple $(3,2,1,1)$ is the signature with the property of minimality.

\section{{\mathversion{bold} Generalized IW-contractions from $\mathrm{so}(3)\oplus A_1$ to $A_{4.1}$}}
\label{SectionOnGenIWcontractionsFromSO3A1ToA41}

In~\cite{Nesterenko&Popovych2006} the contraction  $\mathrm{so}(3)\oplus A_1\to A_{4.1}$ was realized 
as a generalized IW-contraction with the signature $(3,2,1,1)$. 
We show that this signature is minimal in the sense that 
no quadruples with components solely from the set $\{0,1,2\}$ 
can be signatures of a generalized IW-contraction from $\mathrm{so}(3)\oplus A_1$ to $A_{4.1}$. 
Note that the complexification of the contraction $\mathrm{so}(3)\oplus A_1\to A_{4.1}$ is equivalent to 
the contraction $\mathrm{sl}(2,\mathbb C)\oplus\mathfrak g_1\to \mathfrak g_{4.1}$ and, therefore, 
possesses the signature $(1,1,1,0)$, i.e., it is a simple IW-contraction. 
Hence the ground field is essential in this example in contrast to the above considered ones. 

As the contracted algebra $A_{4.1}$ here is the same as in the previous section 
and the contraction is not equivalent to a simple IW-contraction \cite{Huddleston1978}, 
we can use the results (as well as the notation) of the previous section. 
Therefore, it is sufficient to check only the quadruple $(2,1,0,1)$, 
and without loss of generality we can assume that $P$ is equal to the unit matrix.

Suppose that there {\em exists} a generalized IW-contraction $\mathrm{so}(3)\oplus A_1\to A_{4.1}$ with the signature $(2,1,0,1)$ 
and the matrix $U_\varepsilon=AW_\varepsilon$. 
The conditions for existence of the contraction imply, in particular, the following system of algebraic equations 
in entries of the matrix~$A$:
\[
\left(\begin{array}{ccc}b^1_1&b^1_2&b^1_3\\[.3ex]b^2_1&b^2_2&b^2_3\\[.3ex]b^4_1&b^4_2&b^4_3\end{array}\right)Y=
\left(\begin{array}{cc}0&0\\[.3ex]0&1\\[.3ex]0&0\end{array}\right),
\]
where $B=(b^i_j)=A^{-1}$ and 
\[
Y=
\left(\begin{array}{cc}
y^1_1&y^1_2\\[.3ex]
y^2_1&y^2_2\\[.3ex]
y^3_1&y^3_2
\end{array}\right)
:=
\left(\begin{array}{cc}
a^2_2a^3_3-a^3_2a^2_3&a^2_3a^3_4-a^3_3a^2_4\\[.3ex]
a^3_2a^1_3-a^1_2a^3_3&a^3_3a^1_4-a^1_3a^3_4\\[.3ex]
a^1_2a^2_3-a^2_2a^1_3&a^1_3a^2_4-a^2_3a^1_4
\end{array}\right).
\]
We complete the system with zero terms and ``virtual'' equations:
\begin{gather}\label{EqSO3A1A41Grading2101Virtual}
B\left(\begin{array}{cc}
y^1_1&y^1_2\\[.3ex]
y^2_1&y^2_2\\[.3ex]
y^3_1&y^3_2\\[.3ex]
0&0
\end{array}\right)
=\left(\begin{array}{cc}0&0\\[.3ex]0&1\\[.3ex]x_1&x_2\\[.3ex]0&0\end{array}\right), 
\quad\mbox{i.e.,}\quad
\left(\begin{array}{cc}
y^1_1&y^1_2\\[.3ex]
y^2_1&y^2_2\\[.3ex]
y^3_1&y^3_2\\[.3ex]
0&0
\end{array}\right)
=A\left(\begin{array}{cc}0&0\\[.3ex]0&1\\[.3ex]x_1&x_2\\[.3ex]0&0\end{array}\right), 
\end{gather}
where $x_1$ and $x_2$ are some new indeterminates.
Equating the first columns of the left and right hand sides in the last matrix equation, 
we obtain a system which implies, in view of the condition $\det A\ne0$, that $x_1=0$. 
(The fact that the ground field is real is essential here.)
Therefore, $y^1_1=y^2_1=y^3_1=0$, i.e., the tuples $(a^1_2,a^2_2,a^3_2)$ and $(a^1_3,a^2_3,a^3_3)$ are proportional.
Then the analysis of the system obtained by equating the second columns leads to the assertion 
that the entire second and third columns of~$A$ are proportional. 
This contradicts the nonsingularity~of~$A$.

\noprint{
Equating the first columns of the left and right hand sides in the last matrix equation, 
we obtain the system 
\begin{gather*}
x_1a^1_3=a^2_2a^3_3-a^3_2a^2_3,\\
x_1a^2_3=a^3_2a^1_3-a^1_2a^3_3,\\
x_1a^3_3=a^1_2a^2_3-a^2_2a^1_3,\\
x_1a^4_3=0
\end{gather*}
which implies that $x_1=0$ (since otherwise $a^1_3=a^2_3=a^3_3=a^4_3=0$ and, therefore $\det A=0$). 
This means that the tuples $(a^1_2,a^2_2,a^3_2)$ and $(a^1_3,a^2_3,a^3_3)$ are proportional. 
Analogously, equating the second columns results in the system  
\begin{gather*}
a^1_2+x_2a^1_3=a^2_3a^3_4-a^3_3a^2_4,\\
a^2_2+x_2a^2_3=a^3_3a^1_4-a^1_3a^3_4,\\
a^2_2+x_2a^3_3=a^1_3a^2_4-a^2_3a^1_4,\\
a^2_2+x_2a^4_3=0.
\end{gather*}
}

\section{Discussion of technique applied}
\label{SectionDiscussionOfTechniqueApplied}

The proof of Theorem~\ref{TheoremOnNonexistenceOfGenIWContractionsBetween4DimComplexLieAlgebras}
has a number of special features which, when combined, form a technique applicable to a wide range of similar problems.
For this reason we decided to list them below.

\begin{enumerate}\itemsep=0ex
\item
All necessary criteria for general contractions \cite{Burde2005,Burde&Steinhoff1999,Nesterenko&Popovych2006}
do not work for the study of generalized IW-contractions
since the contraction is known to exist and, therefore, the necessary criteria are definitely satisfied.
The problem is to determine whether the contraction can be realized in a special way
and this requires other tools.
\item
There exists a simple criterion stating that a contraction is not equivalent to a generalized IW-contraction
if the contracted algebra admits improper gradings only.
In contrast with the contractions to characteristically nilpotent Lie algebras,
this criterion is not applicable to the algebra $\mathfrak g_1\oplus \mathfrak g_{3.2}$
since the latter has nontrivial diagonal derivations and therefore possesses proper gradings.
\item
In the canonical basis, the algebra $\mathfrak g_1\oplus \mathfrak g_{3.2}$ has a two-dimensional algebra of diagonal derivations.
Therefore, we have to consider a number of different gradings for the contracted algebra.
The study of the gradings aims at resolving a twofold challenge---to obtain possible values of parameter exponents
and to understand the structure of constant components of contraction matrices.
Thus, the structure of derivations of the algebra $\mathfrak g_1\oplus \mathfrak g_{3.2}$ implies
that only signatures of the form $(\beta,\alpha,\alpha,0)$ are admissible.
\item
Further restrictions on parameter exponents follow from the absence of simple IW-contractions
from $2\mathfrak g_{2.1}$ to $\mathfrak g_1\oplus \mathfrak g_{3.2}$.
Up to positive multipliers, any signature associated with a simple IW-contraction consists of zeros and units.
Hence we have the condition $0\ne \alpha \ne \beta \ne 0$.
\item
The matrix~$P$ in the representation~$U_\varepsilon=AW_\varepsilon P$ of the contraction matrix $U_\varepsilon$ is determined
up to changes of basis within graded components and up to automorphisms of the contracted algebra.
Since in the case under consideration the matrix $P$ provides an isomorphism among gradings,
we can set $P$ equal to the unit matrix.
\item
A significant part of subcases for parameter exponents can be ignored as the associated systems of equations for entries of the matrix $A$
are extensions of their counterparts for other subcases
and hence the inconsistency of the former systems is immediate from that of the latter ones.
\item
Using the scaling automorphisms of the contracted (or initial) algebra, we set $\det A=1$ to simplify the entries of the inverse matrix $A^{-1}$.
\item
We consider each tuple of parameter exponents for which the corresponding system of algebraic equations for entries of the matrix $A$ is minimal.
This nonlinear system is represented in a specific form that allows us to apply methods of solving {\em linear} systems of algebraic equations.
In particular, we try, wherever possible, to avoid writing out the entries of the inverse matrix $B=A^{-1}$
in terms of entries of the matrix $A$.
\end{enumerate}

Proving that generalized IW-contractions $2\mathfrak g_{2.1}\to\mathfrak g_{4.1}$ and $\mathrm{so}(3)\oplus A_1\to A_{4.1}$
with nonnegative integer parameter exponents
necessarily include exponents which are not less than three 
(see Sections~\ref{SectionOnGenIWcontractionsFrom2G21ToG41} and~\ref{SectionOnGenIWcontractionsFromSO3A1ToA41})
is also based on the above technique.

\section{Conclusion}\label{SectionConclusion}

The main result of the present paper is important from a number of different points of view.
First of all, it gives the exact value of the lowest dimension
for which some of well-defined contractions are not realized by generalized IW-contractions.
This is the first example of such contractions in dimension less than seven.
Moreover, this is also the first example of nonexistence of generalized IW-contraction
for the case when the contracted algebra admits nontrivial diagonal derivations.
The previous series of examples constructed by Burde \cite{Burde1999,Burde2005} for dimensions
greater than six involve characteristically nilpotent algebras possessing nilpotent derivations only.
The very fact of ending the long-lived illusion of universality of generalized IW-contractions
could be of interest for the physical community.
In this connection it is important to stress that the Lie algebras involved are considerably less exotic
than the characteristically nilpotent algebras and appear, for instance, in general relativity~\cite{Petrov1966}.
Thus, the algebra $2A_{2.1}$ can be easily realized as the Lie algebra of the Lie group generated
simultaneous scalings and translations in two directions.

The complete solution of the problem of characterizing generalized IW-contractions 
of four-dimensional complex (resp.\ real) Lie algebras leads to a number of 
other interesting open problems.

It is now known that all contractions of three-dimensional complex (resp.\ real) Lie algebras
can be realized via generalized IW-contractions \cite{Nesterenko&Popovych2006}
and that this is not true for the dimension four (the present paper) and the dimensions greater than six (\cite{Burde1999,Burde2005}).
Similar results for dimensions one and two are trivial.
The problem of universality of generalized IW-contractions for five- and six-dimensional Lie algebras is still open.
It is expected that for these dimensions the answer and the approach to this problem will be similar to those for the dimension four.

Since generalized IW-contractions are not universal in the whole set of Lie algebras,
the following question is natural and important:
In which classes of Lie algebras closed under contractions any contraction is equivalent to a generalized IW-contraction?
For example, the classes of four- and five-dimensional nilpotent algebras do have this property
\cite{Burde&Nesterenko&Popovych2009,Grunewald&Halloran1988,Nesterenko&Popovych2006}.

Although the total universality of generalized IW-contractions was disproved by counterexamples \cite{Burde1999,Burde2005},
it was conjectured in~\cite{Campoamor-Stursberg2008} after analyzing
the classification of contractions of four-dimensional Lie algebras presented in~\cite{Nesterenko&Popovych2006}
that any contraction of Lie algebras is a composition of generalized IW-contractions.
Examples of \cite{Burde1999,Burde2005} also provide counterexamples for the latter conjecture.
There is a contraction among seven-dimensional characteristically nilpotent Lie algebras with
orbit dimensions differing by 1.
Therefore, this contraction is indecomposable and is not equivalent to a generalized IW-contraction.
Representing general contractions of nilpotent algebras via generalized IW-contractions
is studied in~\cite{Burde&Nesterenko&Popovych2009} at greater length.
One can state a weaker conjecture that any contraction to a Lie algebra possessing  nontrivial gradings
is a composition of generalized IW-contractions.
This conjecture does not contradict already known four- and seven-dimensional examples of contractions
inequivalent to a generalized IW-contraction but it is expected that suitable counterexamples may be found.

The last but not least problem is to find criteria for existence of generalized IW-contractions
which would be different from the simplest one, based on testing whether there are any gradings at all in contracted algebras,
and would be powerful enough for the case when the contracted algebra possesses non-nilpotent derivations.

\subsection*{Acknowledgements}

The authors are grateful to Dietrich Burde, Maryna Nesterenko, Anatoly Nikitin, Artur Sergyeyev and Evelyn Weimar-Woods
for productive and helpful discussions and thank the referee for useful remarks.
The research of ROP was supported by the Austrian Science Fund (FWF), project P20632.

\end{document}